# Thermal diffusivity recovery and defect annealing kinetics of self-ion implanted tungsten probed by insitu Transient Grating Spectroscopy

Abdallah Reza[1*], Guanze He[1], Cody A. Dennett[2], Hongbing Yu[1,3], Kenichiro Mizohata[4], Felix Hofmann[1†]

[1]Department of Engineering Science, University of Oxford, Parks Road, Oxford, OX1 3PJ, UK

[2] Materials Science and Engineering Department, Idaho National Laboratory, Idaho Falls, ID 83415, USA

[3]Canadian Nuclear Laboratories, Chalk River, Ontario, Canada, K0J1J0

[4] Materials Physics, University of Helsinki, P.O. Box 64, 00560 Helsinki, Finland

## Graphical Abstract

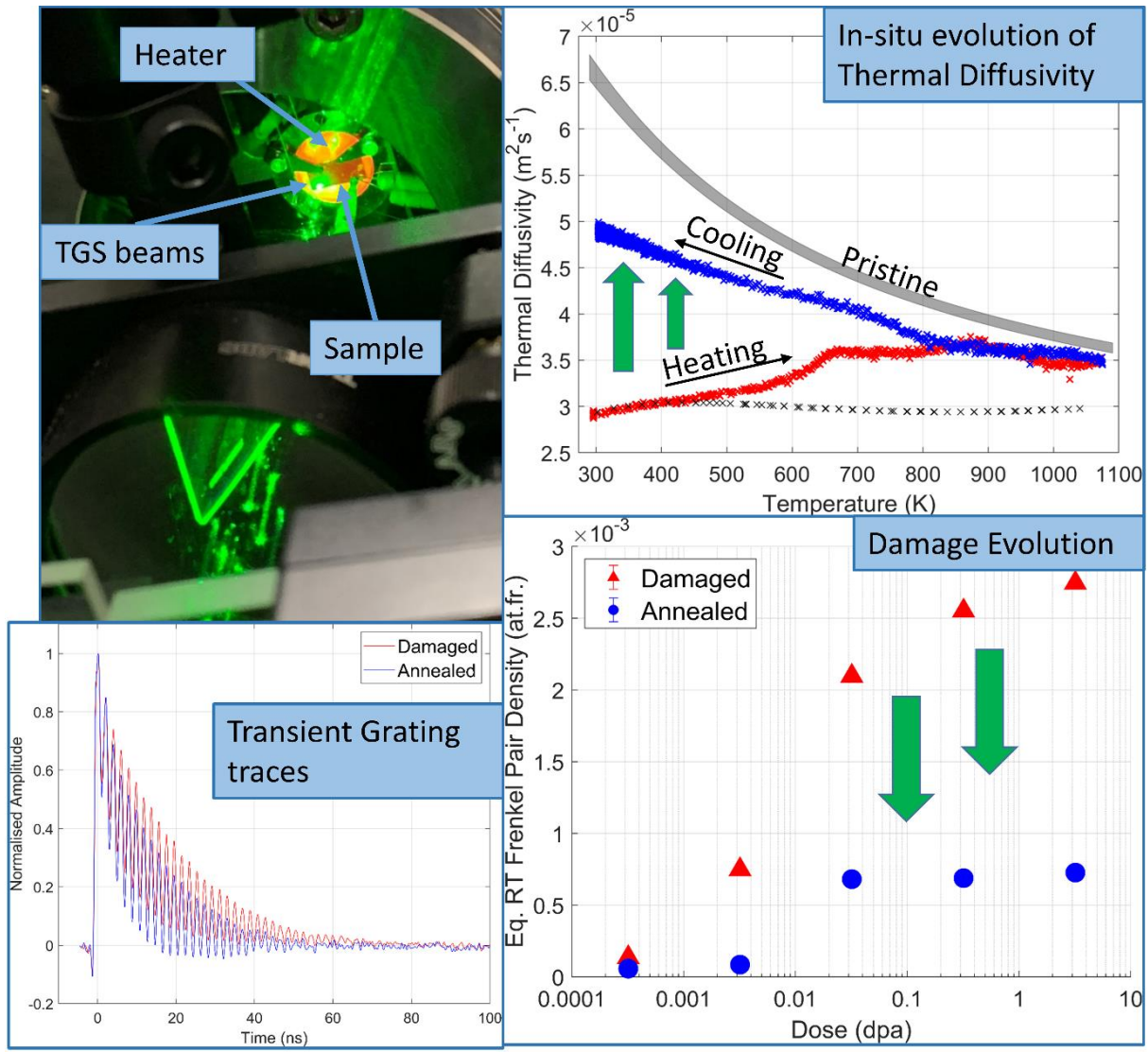

Corresponding author:   *  mohamed.reza@eng.ox.ac.uk          † felix.hofmann@eng.ox.ac.uk

## Abstract

Tungsten is a promising candidate material for plasma-facing armour components in future fusion reactors. A key concern is irradiation-induced degradation of its normally excellent thermal transport properties. In this comprehensive study, thermal diffusivity degradation in ion-implanted tungsten and its evolution from room temperature (RT) to 1073 K is considered. Five samples were exposed to 20 MeV self-ions at RT to achieve damage levels ranging from $3.2 \times 10^{-4}$ to 3.2 displacements per atom (dpa). Transient grating spectroscopy with insitu heating was then used to study thermal diffusivity evolution as a function of temperature. Using a kinetic theory model, an equivalent point defect density is estimated from the measured thermal diffusivity. The results showed a prominent recovery of thermal diffusivity between 450 K and 650 K, which coincides with the onset of mono-vacancy mobility. After 1073 K annealing samples with initial damage of $3.2 \times 10^{-3}$ dpa or less recover close to the pristine value of thermal diffusivity. For doses of $3.2 \times 10^{-2}$ dpa or higher, on the other hand, a residual reduction in thermal diffusivity remains even after 1073 K annealing. Transmission electron microscopy reveals that this is associated with extended, irradiation-induced dislocation structures that are retained after annealing. A sensitivity analysis shows that thermal diffusivity provides an efficient tool for assessing total defect content in tungsten up to 1000 K.

# 1 Introduction

Tungsten and its alloys are frontrunners for use in plasma facing armour components of future fusion reactors [1–4]. This is due to the high thermal conductivity, high melting point, low sputtering yield and low vapour pressure of tungsten [3,5]. In service, armour components must withstand a harsh operating environment with high temperatures, intense neutron irradiation and large heat flux [1,3,6]. Neutron irradiation can alter the structure and chemistry of tungsten, changing its physical properties such as its strength and thermal conductivity [7–10]. Room temperature (RT) thermal conductivity of pure tungsten is known to reduce dramatically with irradiation; up to ~50% once 0.1 displacements per atom (dpa) of displacement damage is reached. Similarly helium injection decreases it by ~55% with just 3000 appm of He ingress [7,9]. In the reactor, a degradation of thermal diffusivity would lead to steeper temperature gradients within components for the same heat load. This would cause (a) higher surface temperatures that could lead to cracking and melting, and (b) larger thermal stresses that would accelerate fatigue [1,11]. A comprehensive understanding of the evolution of tungsten's thermal diffusivity with radiation and temperature is thus essential for the design of safe and efficient future fusion reactors.

In addition to neutron bombardment, tungsten will also be exposed to deuterium and helium that may diffuse into the material from the plasma [4,12–14]. Helium can form micro-bubbles within the matrix and create surface fuzz [15,16] while deuterium creates gas-filled voids and causes surface blistering [14,17]. Neutron bombardment can induce transmutation to form rhenium, osmium and tantalum that reduce the purity of the material [18–20]. Helium is also produced by transmutation, in larger quantities than those entering the matrix from the plasma, and deeper beneath the surface [12]. In addition to transmutation, neutron bombardment can displace lattice atoms creating self-interstitials and leaving behind vacancies. The displaced atoms can go on to knock further atoms off their lattice sites, creating cascades of displacement damage. These defects can elastically interact and evolve to form more complex defect structures that modify material properties such as the thermal conductivity, elastic modulus and others [10,21]. In this study, we focus on the temperature and dose evolution of displacement damage and the associated thermal diffusivity degradation.

High energy self-ion implantation mimics the displacement damage created by fusion neutrons, while side-stepping additional complexities associated with neutron irradiation, such as transmutation. The defect population created by RT ion bombardment spans a range of sizes, from single atom defects to large defect clusters and dislocation loops [22–24]. At low doses the defect size distribution follows a power law with a negative exponent, giving more small defects (point defects and small clusters) and fewer large defects (large clusters, loops and voids) [23,25]. At medium and high doses, the defect population can be more complex as there is significant overlap of displacement damage cascades, which leads to the formation of larger defect structures and dislocation networks [26]. Recent modelling work suggests that at high doses a dense population of small defects persists, in addition to the larger defect structures [27]. Traditional defect imaging techniques such as transmission electron microscopy (TEM) have limitations on the smallest size of defects that can be distinguished, usually ~1.5 nm [28]. Hence most studies have not considered the effects, defects below this size [7,29]. Yet, previous studies suggest that these 'hidden' defects may significantly affect mechanical properties [29] and thermal diffusivity [7].

The effect of irradiation temperature on defect populations in tungsten has been probed extensively in the past, including work on ion-implantation [30–32] and neutron irradiation [33]. RT ion-implantation studies showed TEM visible loops (>1.5 nm), that were vacancy loops. The mechanism of their formation is believed to be from the collapse of regions having a super-saturation of vacancies at the core of displacement cascades. For higher temperature implantations, ~500°C, interstitial loops are also observed at higher doses [32]. However the defect yields at high temperature were lower than those at RT [30]. This is due to vacancy mobility which enables long-range migration, thus reducing the local concentration below the nucleation threshold for loops. Annealing of implanted tungsten to higher temperatures (> 800°C) showed significant increase of loop size, accompanied by a reduction in the number of loops [34]. The observed loops were also reported to be of interstitial type. Temperature plays a key role in the mobility of the different defect structures and thus the damage population that evolves.

Several studies have also been undertaken on post-irradiation annealing, examining the recovery of defects in tungsten. Most of these studies used indirect methods such as electrical resistivity measurements [35–38] and only recently direct TEM observation [34]. These studies considered ex-situ annealing followed by the resistivity measurements / microscopy. Thus far there has been only one study on implantation-induced defects in tungsten, which used TEM with insitu irradiation and heating [24]. Resistivity measurements and positron annihilation lifetime spectroscopy (PALS) have been particularly useful in identifying the recovery stages of specific types of defects and the corresponding temperatures in tungsten [39]. It should also be noted that defect recovery stages differ in single-crystal and polycrystalline tungsten, due to the absence of grain boundaries in single-crystals, which act as sinks for defects [34,37]. The first stage of defect recovery occurs below ~100 K, where free interstitials become mobile and begin to travel around the lattice towards vacancies and other sinks [34]. Between ~100 K and ~520 K is the second stage where interstitials can be released from defect traps such as cascades [40]. The third stage of defect recovery centers around monovacancies, which become mobile at ~520 K [39,40]. A fourth recovery stage at ~800 K in tungsten has been generally attributed to vacancy-impurity complexes and di-vacancies [37]. A fifth stage at ~1100 K is loosely attributed to the formation of voids and the breaking down of defect clusters [34,41]. For the temperature range covered in this study, i.e. RT-1073 K, defect recovery stages 2-4 are applicable.

Although self-ion implantation is an easier alternative to neutron irradiation, it poses a challenge due to the fact that it only produces a thin damaged layer (~µm). Transient grating spectroscopy (TGS) provides us with the unique capability of rapid, non-contact and accurate measurements of thermal diffusivity in micron thick implanted layers [9,42]. Kinetic theory models can utilise these measurements to provide estimates of the underlying defect densities [7,43].

In this study we demonstrate that TGS is able to capture the evolution of thermal diffusivity in thin self-ion implanted tungsten surface layers during insitu annealing (section 3.1). An estimate of the underlying defect densities is obtained from the measured thermal diffusivities using simple kinetic theory point scattering models (section 3.2). The insitu heating capability provides high resolution in temperature, thus making it possible to identify the onset of different defect healing mechanisms as well as estimate the proportion of defects removed. TEM is performed to confirm predictions of microstructural evolution obtained from the thermal diffusivity measurements and defect density estimates (section 3.3).

## 2 Materials and Methods

### 2.1 Sample preparation and implantation

The samples for this study were a subset of those used in [7] and thus their preparation and implantation procedure is only briefly summarised here. Polycrystalline tungsten samples with >99.97 wt% purity procured from Plansee were cut into 10 mm squares (1 mm thick), and annealed for 24 hrs at 1723 K in a vacuum of ~$10^{-5}$ bar. The samples underwent a polishing process that included mechanical, diamond slurry and electro-polishing steps. An electron back scatter diffraction (EBSD) map of the $3.2 \times 10^{-2}$ dpa sample showing the average grain sizes is given in appendix A.

Ion implantations were carried out with 20 MeV $^{184}W^{5+}$ ions, using a 5 MeV tandem accelerator at the Helsinki Accelerator Laboratory [44]. The beam spot size used was ~5 mm, and raster scanning was utilised over a 15x15 mm$^2$ region to obtain a uniform implantation profile. Five dose levels were considered: $3.2 \times 10^{-4}$, $3.2 \times 10^{-3}$, $3.2 \times 10^{-2}$, $3.2 \times 10^{-1}$ and 3.2 dpa. The lowest and highest dose samples were implanted with flux densities of $6.2 \times 10^8$ and $1.1 \times 10^{11}$ ions/cm$^2$/s respectively. The rest were implanted at ~$3.1 – 5.0 \times 10^{10}$ ions/cm$^2$/s. All implantations were "nominally" at room temperature. We estimate that the maximum temperature rise for the highest dose sample is no more than 100 K, and less for lower doses. Fig. 1(b) shows a normalised damage profile of the samples, obtained from SRIM [45]. For the SRIM calculations a threshold displacement energy of 68 eV was used for tungsten [46], with the 'Quick Kinchin-Pease' calculation model. The implanted samples were sectioned into four equally-sized quartets, using a diamond tipped fast saw, to enable separate analysis whilst having the same implantation history. The as-implanted samples have been considered in several separate studies concerned with irradiation induced changes in lattice strain [47], plastic deformation behaviour [48], defect structure formation [49] and thermal diffusivity [7].

### 2.2 Thermal diffusivity measurements with insitu annealing

**Thermal Diffusivity**

Thermal diffusivity measurements were carried out using the transient grating spectroscopy (TGS) method [50–53]. Details of the setup used can be found in [54]. Two excitation pulses (532 nm wavelength, 0.5 ns pulse duration, 1 kHz repetition rate, 1.5 μJ average pulse energy at sample) were crossed at the sample surface at a well-defined angle, to create a spatially periodic intensity pattern, with a well-defined wavelength. The sample surface partially absorbs the light, creating a thermal grating. Rapid thermal expansion creates a displacement grating and launches two counter-propagating surface acoustic waves (SAWs). The thermal and displacement grating are referred to as the 'Transient Grating'. After the excitation pulse, heat flows from the peaks to the troughs of the temperature grating as well as into the bulk, and thus the temperature grating decays. The rate of this decay depends on the thermal diffusivity of the sample's surface region [51] . The SAWs introduce a temporal oscillation of the surface displacement grating that depends on the elastic properties of the sample [10,55] .

The rate of decay of the transient grating is probed by diffracting a 'probe' laser beam (continuous wave, 559.5 nm wavelength) from it. The intensity of the diffracted beam, monitored by an avalanche photodiode, decays with a time constant that contains the thermal diffusivity signature of the sample. The probe beam is heterodyned with a reference beam for increased signal intensity [50,53,54,56,57]

Fig. 1(a) shows sample traces for the 3.2 dpa sample, before and after annealing. The unannealed state exhibits a slower signal decay, indicating a lower thermal diffusivity. The oscillations in the signals are due to the counter-propagating SAWs [58].

The wavelength of the transient grating, i.e. the TGS wavelength used in this study, was set at $\lambda$ = 5.116 ± 0.001 μm. This sets the thickness of the sample layer within which thermal diffusivity is probed, $\sim \lambda/\pi$, at ~1.6 μm [51]. From Fig. 1(b) we see that this probing depth is fully contained within the thickness of the surface layer damaged by self-ions. The mean probe beam power was ~22 mW (at the sample), which was the total for both probe beams and reference beams as the dual-heterodyne technique was used [53]. The average power of the excitation beams at the sample was ~1.5 mW (1.5 μJ pulses at 1 kHz). The average excitation and probe power absorbed by the sample was ~0.75 mW and ~11 mW respectively, consistent with a sample reflectivity of ~50 %. The laser spot size at the sample was ~90 μm and ~140 μm for the probe and excitation respectively (1/e$^2$ level).

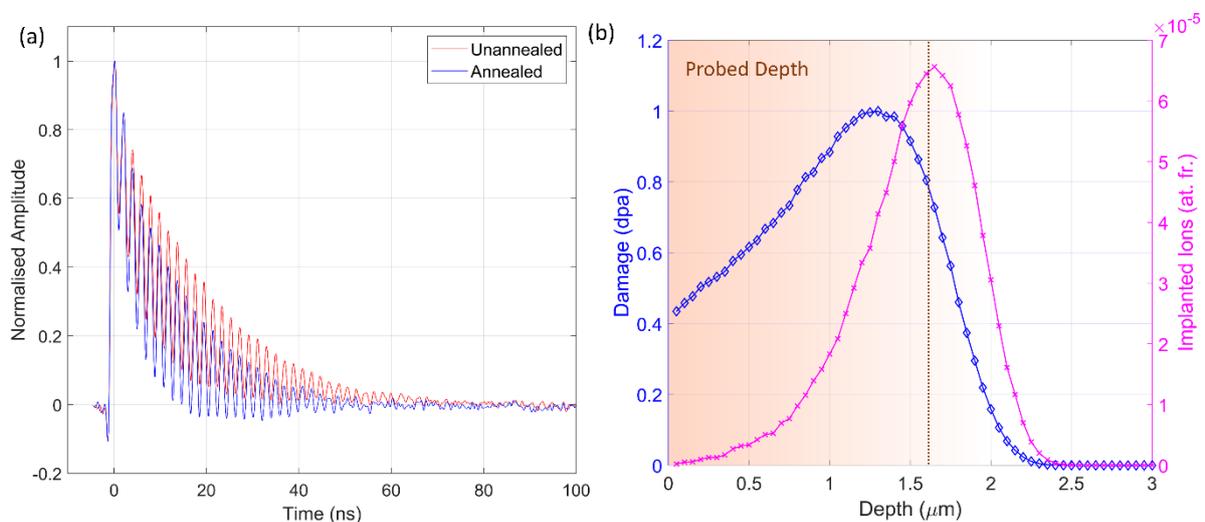

Figure 1. (a) Sample TGS traces for the annealed and unannealed 3.2 dpa sample. (b) normalised damage and implantation profiles predicted by SRIM. The shaded background in (b) indicates the TGS probing depth. The dashed vertical line in (b) shows the $\sim \lambda/\pi$ probing depth for the thermal diffusivity measurement.

**Insitu annealing**

A high temperature resistive button heater was used for insitu annealing (Heatwave labs 101491, 0.5 inch diameter). It is held in a refractory metal heat shield and uses clips (Fig. 2(c)) to press the sample to the heater surface during TGS measurements (Fig. 2 (a)). The heater is rated to 1473 K but was tested only to 1073 K as this was sufficient for this study. Ceramic spacers, shown in Fig. 2(c) are used between the sample clips and the sample surface to prevent heat loss to the clips. A K-type thermocouple placed on the rear surface of the heater, within the heat-shield, is used for temperature control. For the sample temperature reading, a second K-type thermocouple was clamped on to the sample surface using one of the sample clips and a ceramic spacer for thermal and electrical insulation. Fig. 2(c) shows the heater assembly with a tungsten calibration sample in place. As tungsten readily oxidises at elevated temperatures (> ~700 K) [11,59], a vacuum of 10$^{-5}$ mbar or greater was used in this study, provided by a turbo-molecular pump and a diaphragm backing pump. To maintain the alignment of the sample with respect to the laser beams while connected to the pumps, counterbalanced vacuum bellows were used.

To control the heater, a proportional-integral-derivative (PID) controller (Omega Platinum CN8DPt) along with a Darlington transistor was used. Temperature readout and the TGS measurements were automated via MATLAB [60]. For all samples a uniform heating/cooling rate of ~13 K/min (RT - 1073 K in 1 hr) was used. Since no active cooling was implemented, below ~500 K the rate of cooling was less than 13 K/min. The samples were heated to 473 K, 673 K, 873 K and 1073 K, and cooled to RT after each ramp. They were held at the maximum temperature for 2 minutes in each cycle. TGS data was recorded at 20 second intervals (corresponding to a temperature range of 4 K at this heating rate) throughout the heating and cooling segment of each cycle.

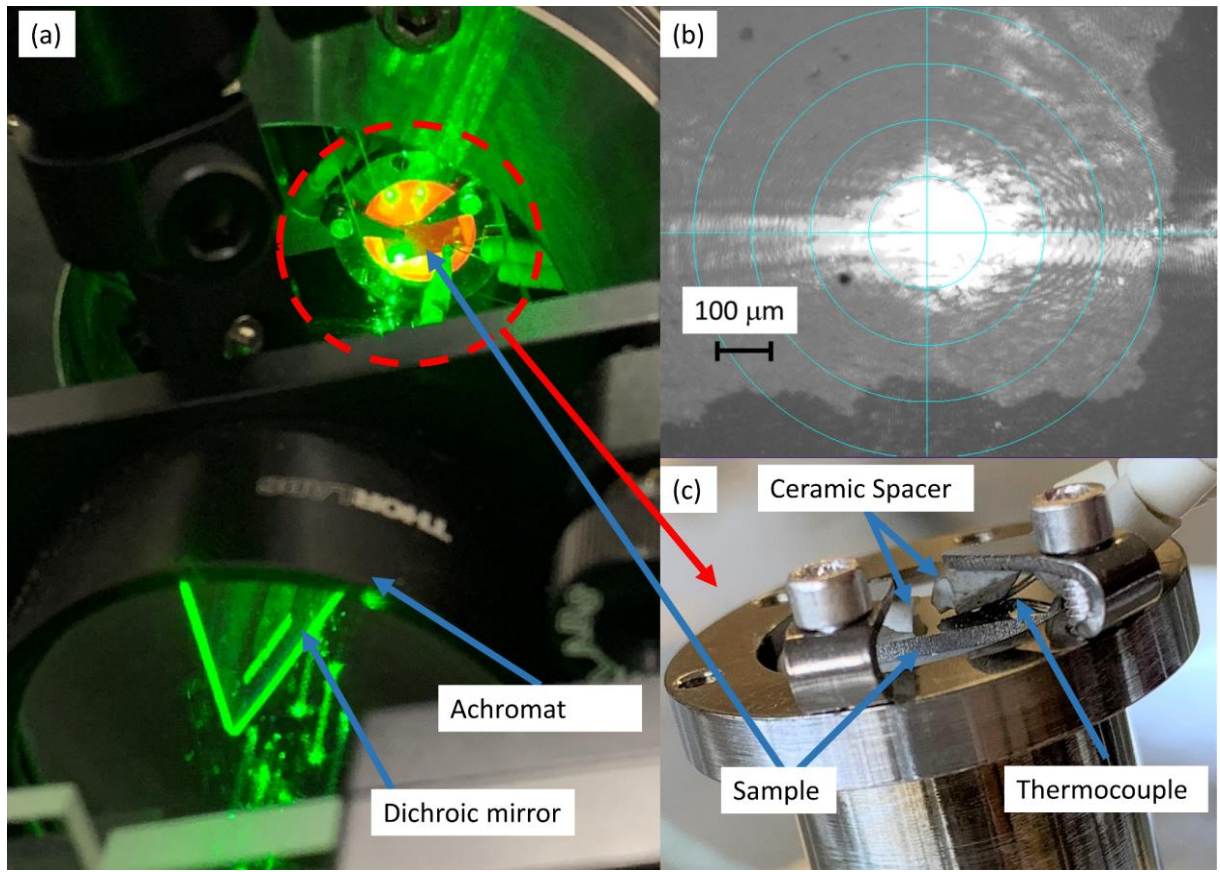

Figure 2. (a) The insitu heater with the sample marked and the laser beams visible on the sample and the achromatic doublet. This image was taken with the heater at ~873 K. (b) View of the sample showing the excitation and probing spots through the in-line microscope. (c) Close-up view of the heater showing the sample, spacers and thermocouple.

## 2.3 Transmission Electron Microscopy

TEM cross-section lamella were lifted out from the as-implanted $3.2 \times 10^{-3}$ dpa and annealed $3.2 \times 10^{-2}$ dpa sample using a Zeiss NVision 40 FIB-SEM fitted with a gallium source. After insitu lift-out, they were subsequently thinned to electron transparency using 30 keV Ga ions and beam currents from 700 pA to 80 pA. A final step of surface cleaning used 5 keV 200 pA Ga ions to remove as much as possible of the FIB induced surface damage. TEM investigation was carried out using a Jeol 2100 TEM with LaB$_6$ source, operating at 200 keV. The TEM specimens were tilted close to weak-beam conditions suitable for radiation defect characterisation.

# 3 Results
## 3.1 Thermal diffusivity

First we examine the thermal diffusivity evolution of the different samples for the various annealing stages from RT to 1073 K (Fig. 3). Data at each measurement point was averaged over 5000 laser excitations to increase accuracy. Sample TGS traces for the 3.2 dpa sample, before and after annealing, are shown in Fig. 1(a). The trace after annealing decreases faster, indicating a higher thermal diffusivity after annealing. The traces were fitted using custom MATLAB [60] scripts, taking into account the thermal diffusivity and SAW contributions to the signal (see eqn. B.1 in appendix B for the fitting equation). A pristine tungsten sample was also measured as a reference and its data plotted as a grey shaded area in Fig. 3. The upper and lower bounds of the region correspond to one standard deviation, after averaging the measurements over 20 K temperature intervals. Data for a hypothetical case where no annealing occurs, for the 3.2 dpa sample, is given by the black cross markers in Fig. 3(e). This was obtained by predicting the RT defect density using the model detailed in section 3.2 and keeping it constant throughout. This is discussed in more detail in section 4.

For the lowest dose sample (3.2 x 10$^{-4}$ dpa, Fig. 3(a)) the unannealed RT thermal diffusivity is $\sim 6.0 \times 10^{-5}\, m^2 s^{-1}$; ~10% lower than the pristine value. After the 673 K anneal, it recovers to the pristine value, suggesting significant defect removal. For the as-implanted 3.2 x 10$^{-3}$ dpa sample, as seen at the start of the 473 K anneal in Fig. 3(b), the RT thermal diffusivity is $\sim 5.0 \times 10^{-5}\, m^2 s^{-1}$. It is observed to be the same after the 473 K anneal, suggesting negligible defect evolution in that temperature range. However, the RT thermal diffusivity after the 673 K anneal is $\sim 5.8 \times 10^{-5}\, m^2 s^{-1}$, indicating that significant annealing has taken place. The 873 K and 1073 K anneals further increase the RT thermal diffusivity to 6.0 and $6.4 \times 10^{-5}\, m^2 s^{-1}$ respectively, thus approaching the unimplanted value.

Likewise, in the 3.2 x 10$^{-2}$ dpa sample (Fig. 3(c)), the 473 K anneal does not alter the RT thermal diffusivity. The 673 K anneal however increases the RT thermal diffusivity from 3.5 to $4.6 \times 10^{-5}\, m^2 s^{-1}$. Subsequent anneals continue to increase the thermal diffusivity up to $5.0 \times 10^{-5}\, m^2 s^{-1}$ after the 1073 K anneal. For the higher doses, i.e. the 3.2 x 10$^{-1}$ dpa and 3.2 dpa, the annealing behaviour is similar to the 3.2 x 10$^{-2}$ dpa sample, where a RT thermal diffusivity of $5.0 \times 10^{-5}\, m^2 s^{-1}$ is reached after the 1073 K anneal, significantly less than the pristine sample value.

Fig. 4 provides a summary of the RT thermal diffusivities measured after the different annealing cycles for all samples. For the higher dose samples ($\geq$ 3.2 x 10$^{-2}$ dpa) the thermal diffusivity only recovers to $5.0 \times 10^{-5}\, m^2 s^{-1}$ after the 1073 K annealing cycle. In the lowest dose sample, complete recovery is seen, returning values close to the book value for pristine tungsten [61]. This suggests that some defects are retained in the high dose samples even at 1073 K. Across all samples, the 673 K annealing cycle consistently gave the most significant recovery of thermal diffusivity. An interesting point to note is the fact that the unannealed 3.2 x 10$^{-3}$ dpa sample and annealed higher dose samples have similar RT thermal diffusivity. In addition, it is worth noting that the RT thermal diffusivity after the 673 K anneal decreases with increasing dose, similar to the RT thermal diffusivity in the as-implanted samples before annealing.

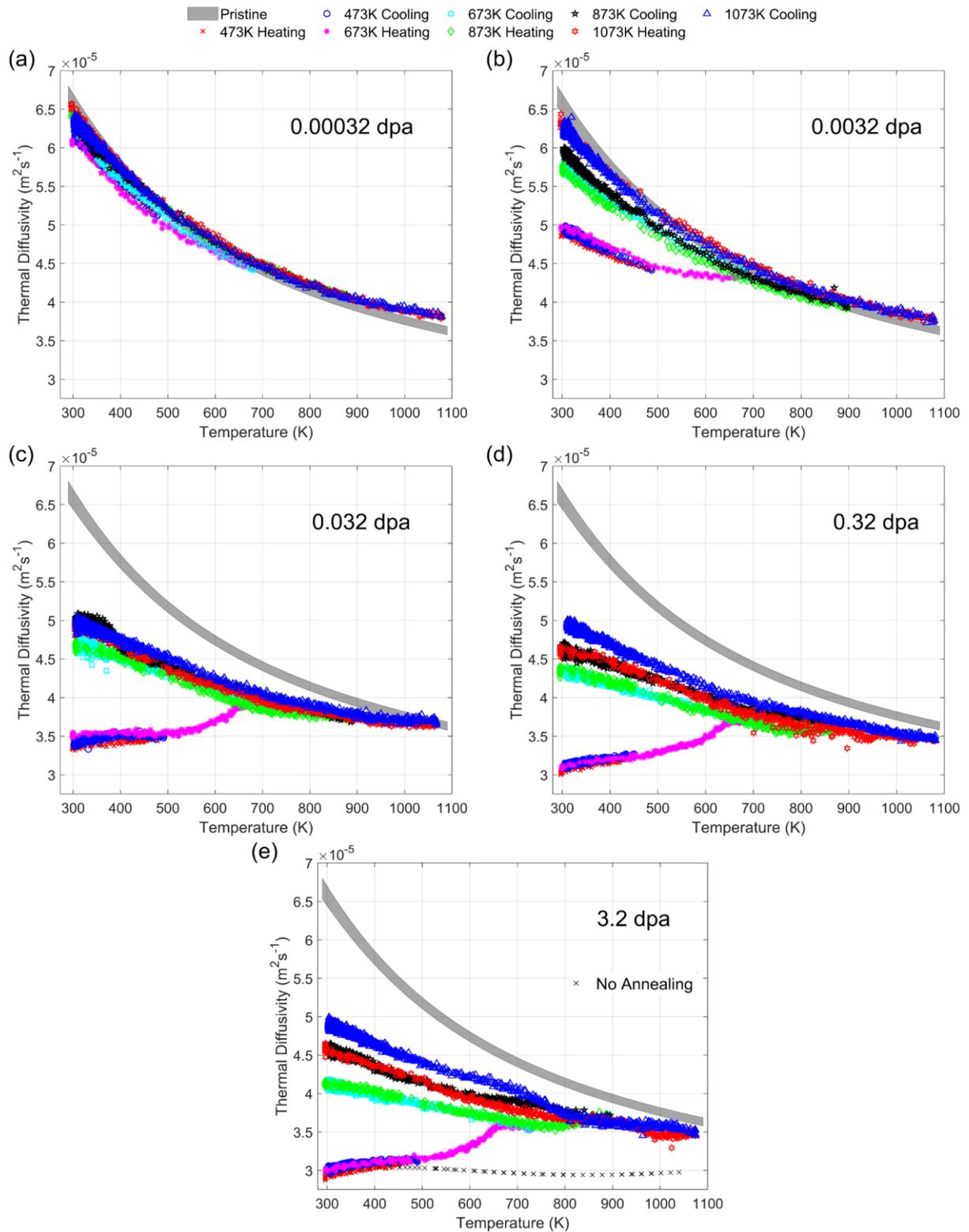

Figure 3. Thermal diffusivity of the implanted samples measured during the various annealing stages. Each measurement is an average of 5000 laser excitations. The thermal diffusivity measured for pristine tungsten is shown as a grey band. The bounds of the grey region correspond to one standard deviation after averaging over 20 K intervals. Calculated data for a hypothetical case where no annealing occurs in the 3.2 dpa sample, is shown by black crosses in (e).

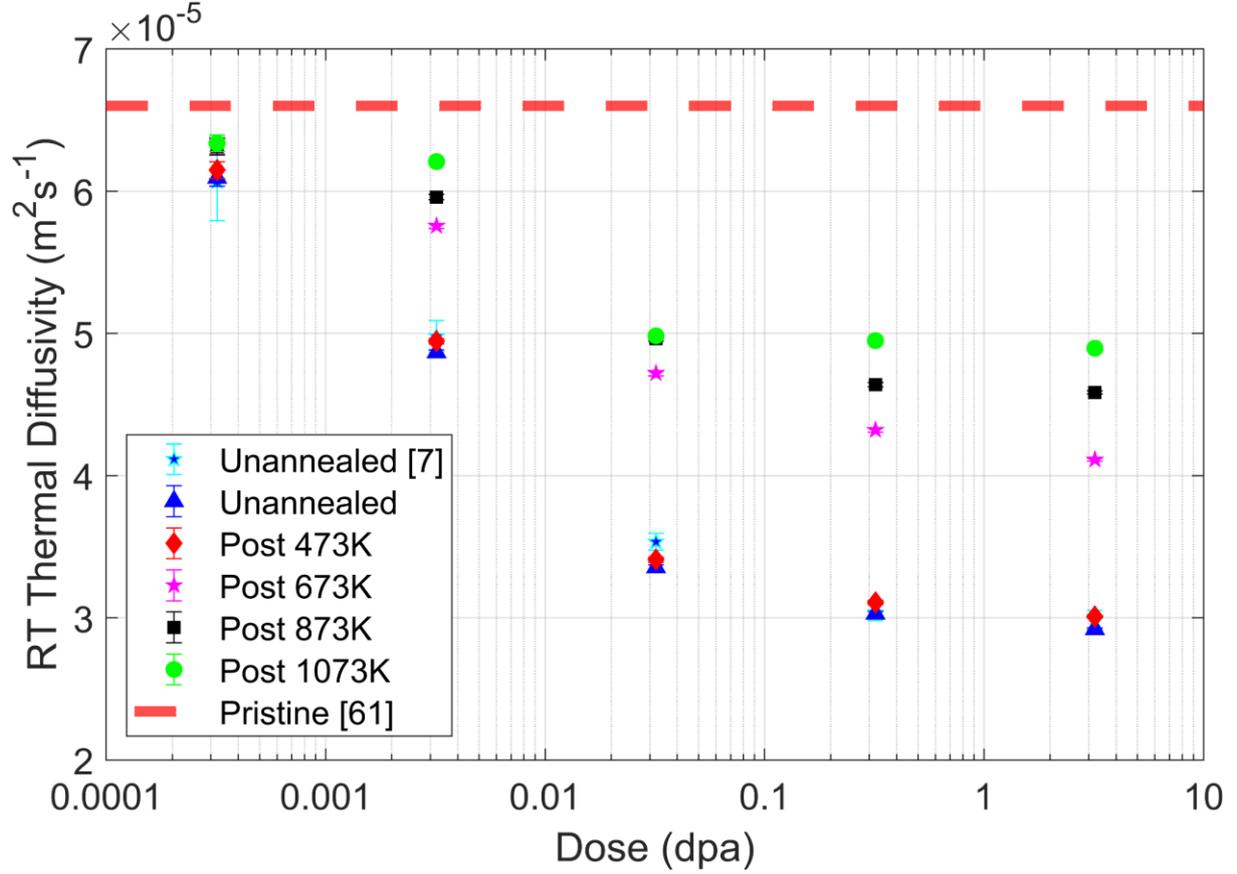

Figure 4. Summary of the RT thermal diffusivity measurements after the different annealing stages for all samples. Also shown are RT thermal diffusivity measurements for RT self-ion implanted tungsten from a previous study [7]. The book-value for the RT thermal diffusivity of pristine tungsten is shown as a dashed red line [61].

## 3.2 Defect annealing kinetics from thermal diffusivity measurements

Above the Debye temperature (312 K for tungsten [62]) electrons are the main carriers of heat. Increased scattering of conduction electrons due to the presence of implantation-induced crystal defects decreases the thermal diffusivity. The present samples are of high purity and thermal diffusivity measurements after annealing and before ion implantation confirmed a thermal diffusivity corresponding to pristine tungsten. The samples are also large grained, with grain sizes on the order of tens of microns (see Fig. A.1 in appendix A for grain maps), and hence grain boundary scattering is negligible [63]. Hence, we can unambiguously attribute the observed reduction in thermal diffusivity to increased electron scattering from crystal defects induced by the ion implantation. The subsequent recovery of thermal diffusivity upon annealing hints at the removal of defects. In this section we examine the evolution of the defect population, by using our thermal diffusivity measurements to estimate the underlying defect densities.

### 3.2.1 Formulation

The electronic thermal conductivity, $\kappa_e$, can be expressed as a function of the electronic heat capacity $C_e$, fermi velocity $v_F$ and electron scattering time $\tau_e$ as follows [64]:

$$\kappa_e = \tfrac{1}{3} C_e v_F^2 \tau_e. \qquad (1)$$

The electron scattering time for pristine tungsten can be expressed as

$$\tau_e = (\sigma_0 + \sigma_1 T + \sigma_2 T^2)^{-1}. \tag{2}$$

where $\sigma_0$ is the temperature-independent scattering contribution from impurities (~0 for pure tungsten), $\sigma_1 T$ is the electron-phonon scattering term that scales linearly with temperature and $\sigma_2 T^2$ is the electron-electron scattering contribution. Their values are determined from the temperature dependence of the thermal diffusivity of pure tungsten [61] and are given in table 3.1.

Considering the self-interstitials and vacancies created by the implantation as impurities in the pure tungsten matrix, the electron scattering time at an impurity of type 'm' is given by

$$\tau_{e,m} = (\sigma_{0,m} + \sigma_1 T + \sigma_2 T^2)^{-1}. \tag{3}$$

Here $\sigma_{0,m}$ is the scattering contribution from the impurity [65,66]. For the defect densities considered in this study, the effect of defects on electron-phonon and electron-electron scattering rates is small, and hence $\sigma_1$ and $\sigma_2$ were assumed to be constant.

The thermal conductivity for the matrix containing defects or impurities, with atomic fraction $c_m$ and scattering time $\tau_{e,m}$ is then given by [9] as

$$\kappa_e = \frac{1}{3} C_e v_F^2 \left( \sum_m \frac{c_m}{\tau_{e,m}} + (1 - \sum_m c_m) \frac{1}{\tau_e} \right)^{-1}. \tag{4}$$

Due to the complexity of the irradiation-induced defect population, in this study we represent the damage caused by irradiation in terms of an equivalent Frenkel pair population. I.e. an interstitial loop with 'x' atoms in the matrix and 'x' individual interstitials in the matrix are treated equally. It should be noted that many types of "collective defects" have been shown to exist, as point defects cluster into different structures. Considering an equivalent point defect density is a necessary simplification and allows a reasonable first order estimate of the underlying defect number density. From here on in this study, the defect number density will be referred to as 'eq. Frenkel pairs' or just as 'defect density'. Equal numbers of interstitials and vacancies are presumed. While loss of interstitials to free surfaces is a significant issue in thin foil TEM samples (~10s of nm) [67], the implanted layer considered in this study was over 2 µm thick, so interstitial loss to the surface or grain boundaries is not expected to play an important role. These assumptions are required to reduce the complexity of the kinetic theory model used for the defect density prediction. A previous study using the same formulation showed good agreement between the predicted defect densities and those from TEM and molecular dynamics (MD) [7]. It should be noted that in order to fully capture the entire effect of irradiation induced damage, the environment of the defects and potentially differing interstitial and vacancy concentrations would need to be considered. This would require additional, detailed knowledge of the dose and temperature dependence of defect evolution, and as such these refinements are not included in the present defect prediction formulation.

Given that the thermal diffusivity can be written in terms of the thermal conductivity $\kappa_e$ and heat capacity $C_P$ as

$$\alpha = \frac{\kappa_e}{C_P}, \tag{5}$$

the expression in eqn. 4 can be used to write the eq. Frenkel pair density $c_{FP}$ as

$$c_{FP} = c_v = c_i = \left[\frac{\frac{\tau_e C_e v_F^2}{3\rho C_P \alpha} - 1}{\tau_e(\sigma'_i + \sigma'_v) - 2}\right], \tag{6}$$

where $\sigma'_i = 1/\tau_{e,i}$ and $\sigma'_v = 1/\tau_{e,v}$ are the scattering rates at interstitial and vacancy sites, respectively, obtained from [9], $\rho$ is the mass density, and $C_P$ is the isobaric heat capacity [68]. The values and/or sources of the parameters used are listed in table 3.1.

Table 3.1 Values and/or sources of parameters used.

| Parameter | Value | Source |
|---|---|---|
| $\sigma_0$ | 4.4 x 10$^{-14}$ s$^{-1}$ | Calculated from [61] |
| $\sigma_1$ | 1.35 x 10$^{11}$ s$^{-1}$ K$^{-1}$ | Calculated from [61] |
| $\sigma_2$ | 8.42 x 10$^7$ s$^{-1}$ K$^{-2}$ | Calculated from [61] |
| $\sigma_{0,i}$ | 16.5 fs$^{-1}$ | Calculated from [69] |
| $\sigma_{0,v}$ | 5.8 fs$^{-1}$ | Calculated from [69] |
| $C_e$ | 87.4T Jm$^{-3}$K$^{-1}$ | [70] |
| $C_P$ | 21.868372 + 8.068661 x 10$^{-3}$T − 3.756196 x 10$^{-6}$T$^2$ + 1.075862 x 10$^{-9}$T$^3$ +1.406637 x 10$^4$T$^{-2}$ | [68] |
| $L$ | 3.2 x 10$^{-8}$ V$^2$ K$^{-2}$ | [71] |
| $v_F$ | 9.50 Å fs$^{-1}$ | [70] |

Inserting the measured value of the thermal diffusivity into eqn. 5 and using literature values for the remaining parameters as detailed in Table 3.1, would give the eq. Frenkel pair density, as seen in [7]. This formulation essentially compares the measured thermal diffusivity to the diffusivity of pure tungsten predicted by the kinetic theory model. It assumes that the behaviour of pristine tungsten exactly matches the model. Hence any minor deviation of pristine tungsten from the perfect kinetic theory model becomes falsely attributed to irradiation induced defects.

A more robust approach is to extract the eq. Frenkel pair density, based on the thermal diffusivity degradation, which is the difference between the measured value for the implanted sample at a given temperature, and that of an unimplanted tungsten reference sample at the same temperature. The implanted and reference samples considered here had the same processing history apart from the ion implantation. The modified expression for the eq. Frenkel pair density is then given by

$$c_{FP} = c_v = c_i = \left[\frac{\Delta\alpha}{[\tau_e(\sigma'_i + \sigma'_v) - 2]\left[\frac{\tau_e C_e v_F^2}{3C_P} - \Delta\alpha\right]}\right], \tag{7}$$

where $\Delta\alpha$ is the thermal diffusivity difference given by

$$\Delta\alpha = \alpha_{pure} - \alpha_{implanted}, \tag{8}$$

and $\alpha_{implanted}$ is the measured value for the implanted sample at a given temperature and $\alpha_{pure}$ is that of an unimplanted tungsten reference sample at the same temperature. It should be noted that $C_e$, $C_P$, $\tau_e$, $\sigma'_i$ and $\sigma'_v$ are all temperature dependent.

The values for $\sigma'_v$ and $\sigma'_i$ are obtained from electrical resistivity data on resistivity per interstitial and vacancies respectively [69] using the method detailed in [7,9]. The Wiedemann-Franz law is used to convert resistivity estimates to thermal conductivity ones, using the Lorenz ratio '$L$' given in table 3.1. $C_e$ and $v_F$ are obtained from [70] and $\tau_e$ is obtained from fitting the thermal diffusivity data for pure tungsten (see eqn. 2 for expression and table 3.1 for values). $C_P$ is calculated using an analytical expression from [68,72]. See appendix B for the uncertainty calculation and table 3.1 for the values used.

### 3.2.2 Results

Plots of the thermal diffusivity difference (difference between measured thermal diffusivity of the relevant sample during the annealing cycles and the thermal diffusivity of a pristine sample from the same batch of material measured at the same temperature) and estimated defect density for the 3.2 x 10$^{-3}$ dpa and 3.2 x 10$^{-1}$ dpa samples are shown in Fig. 5. Plots for the remaining samples are provided in Figs. A.2-A.4 in appendix A.

For the 3.2 x 10$^{-3}$ dpa sample the RT defect density before heating to 473 K is $7.3 \pm 0.1 \times 10^{-4}$ atomic fraction (at.fr.). Upon cooling, it is $7.0 \pm 0.1 \times 10^{-4}$ at.fr.. This hints at the removal of a small proportion of the defects present. After annealing to 673 K, the RT defect density is $3.0 \pm 0.1 \times 10^{-4}$ at.fr., a 57% reduction, implying significant removal of defects. Further annealing to 873 K decreases the RT defect density to $2.0 \pm 0.1 \times 10^{-4}$ at.fr. There is a slight discontinuity between the end of the 873 K anneal and the start of the 1073 K anneal, the cause of which is not clear. It is possible that a shift in the scan location on the sample led to this. After the 1073 K anneal, the defect density is $0.7 \pm 0.1 \times 10^{-4}$ at. fr.. This, and the fact that the thermal diffusivity has recovered implies almost complete removal of the irradiation induced defects.

The 3.2 x 10$^{-1}$ dpa sample has an unannealed RT defect density of $2.55 \pm 0.02 \times 10^{-3}$ at.fr., only a little over three times greater than the 3.2 x 10$^{-3}$ dpa sample, despite a 100 times greater damage dose. The 473 K anneal reduces it slightly to $2.45 \pm 0.02 \times 10^{-3}$ at.fr. The 673 K anneal causes a drastic removal of defects, giving a RT defect density of $1.10 \pm 0.02 \times 10^{-3}$ at.fr., a more than 50% reduction. The 873 K and 1073 K anneals further decrease the RT defect density to $0.92 \pm 0.02 \times 10^{-3}$ and $0.71 \pm 0.02 \times 10^{-3}$ at.fr. respectively.

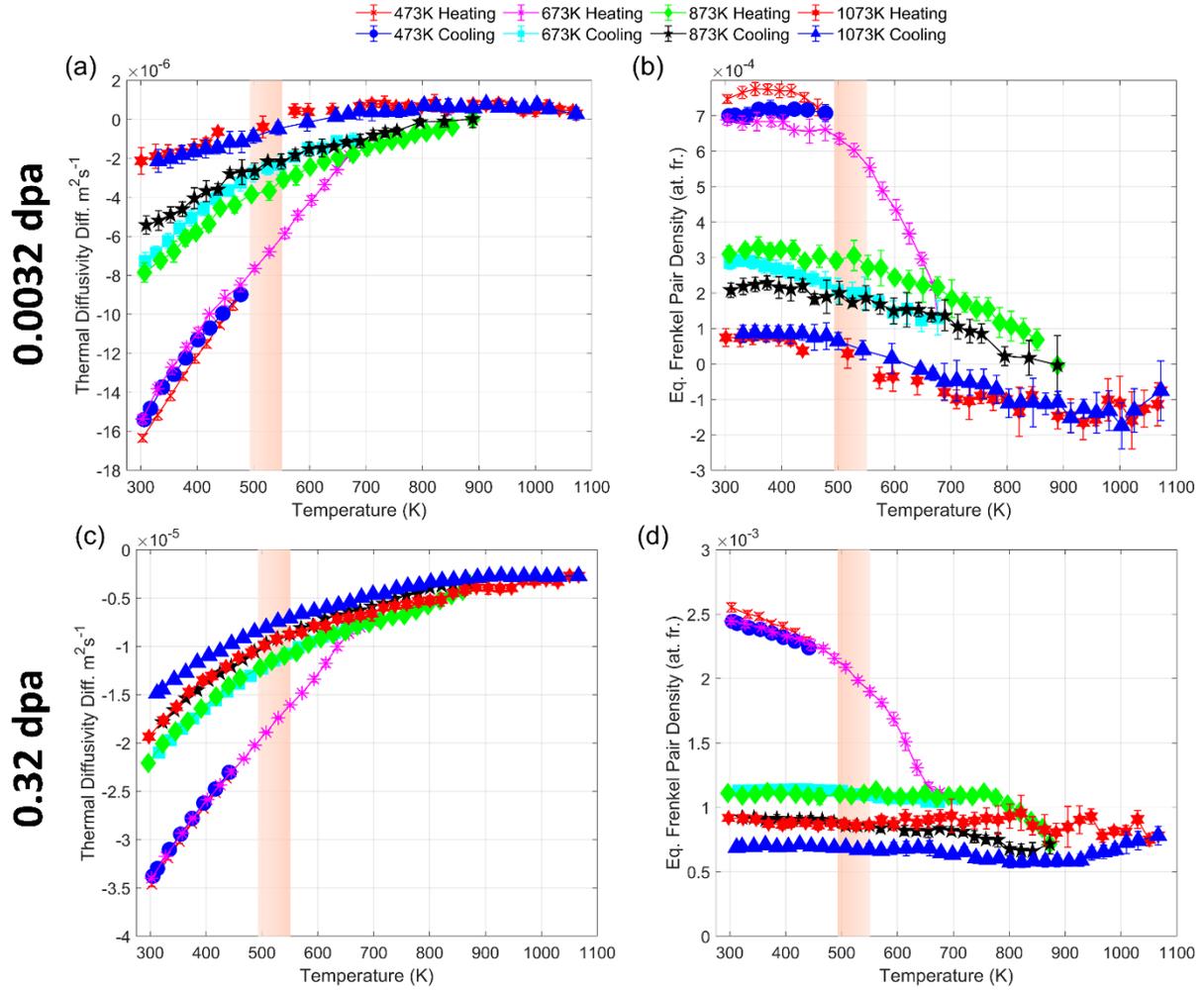

Figure 5. Thermal diffusivity degradation (a), (c) and equivalent Frenkel pair density estimates (b), (d), for the 3.2 x $10^{-3}$ dpa and 3.2 x $10^{-1}$ dpa samples. The data was binned into 20 K wide bins. The shaded region indicates the temperature range of mono-vacancy mobility [73].

Fig. 6 shows RT defect density estimates for all the damage levels considered in this study following the different annealing cycles. The highest dose sample (3.2 dpa) behaves similar to the 3.2 x $10^{-1}$ dpa sample. This is expected as a saturation of defect density has been observed beyond 0.1 dpa in self-ion implanted tungsten [7,24]. It is interesting to note that for the damage levels of 3.2 x $10^{-2}$ dpa and above, the RT defect densities after the 1073 K anneal (green markers in Fig. 6) are surprisingly similar, around $\sim 0.71 \times 10^{-3}$ at.fr.

A common trend across all samples is the significant (~50%) defect removal in the 673 K anneal. Interestingly, studies have shown that while interstitials are mobile well below RT, mono-vacancies; the smallest and most common vacancy-type defects, become mobile in the 500-600 K temperature interval [34,39,73]. Hence we hypothesize that this large reduction in defects is triggered by the onset of single-vacancy mobility, making it possible for vacancies to move to sinks, combine into larger defects, or recombine with self-interstitials.

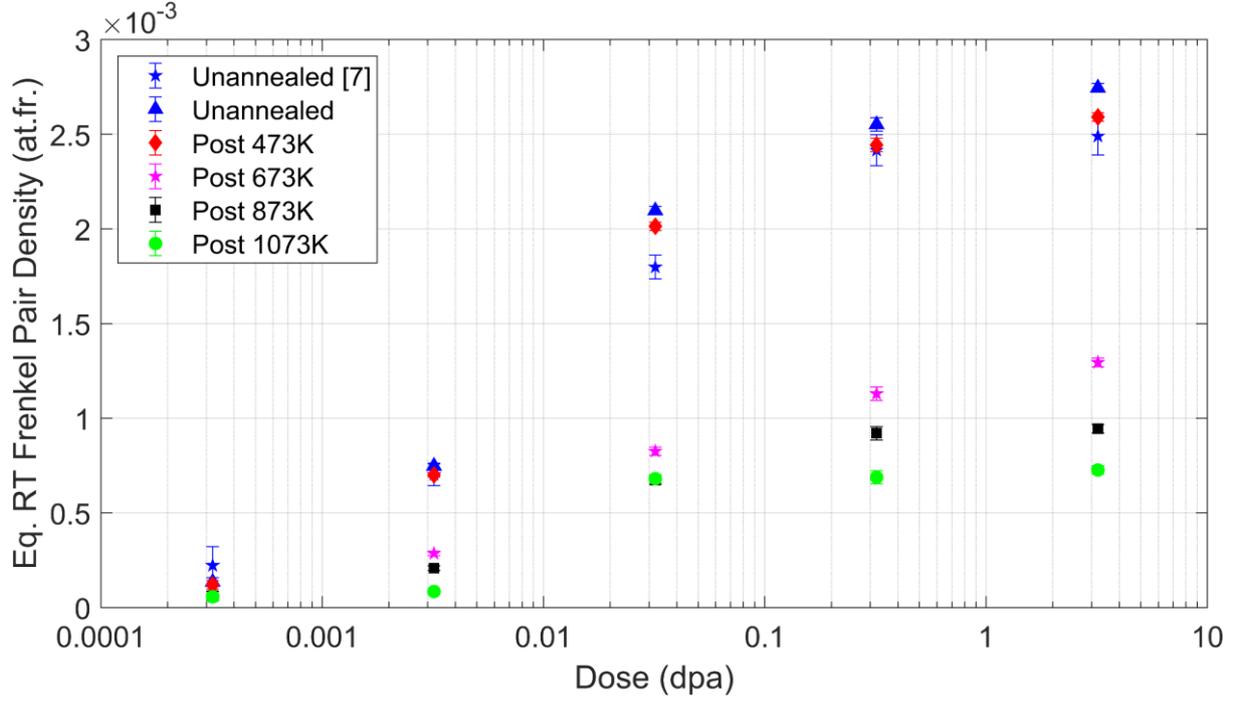

Figure 6. Summary of the RT equivalent Frenkel pair estimates after the various annealing stages for all samples. Also given in the plot are spatially averaged RT TGS defect measurements for self-ion implanted tungsten from a previous study [7].

At higher temperatures in the lower dose samples (e.g. 3.2 x 10$^{-3}$ dpa in Fig. 5(b), and 3.2 x 10$^{-4}$ dpa in Fig. A.2 and 3.2 x 10$^{-2}$ dpa in Fig. A.3 in appendix A) defect density decreases with heating, only to increase again upon cooling. This effect is believed to not be physical but rather the combined effect of the reduced sensitivity of the thermal diffusivity to defects at high temperatures, the inherent uncertainty in the measurement and systematic error. This effect is seen to be amplified with decreasing dose, i.e. more prominent, in the 3.2 x 10$^{-4}$ dpa sample, and for the higher temperature anneals where the previous anneals have further reduced the defect densities. The defect densities in general in the 3.2 x 10$^{-4}$, 3.2 x 10$^{-3}$ and 3.2 x 10$^{-2}$ dpa samples are quite low and this effect corresponds to a significant portion of the total defect density, making it more prominent. Considering systematic errors, the thermal diffusivity measurements are regarded to be accurate up to ~2 x 10$^{-6}$ m$^2$s$^{-1}$ which is ~3% for pristine tungsten at RT, in line with uncertainties expected for TGS measurements.

The reduced sensitivity at high temperatures of the thermal diffusivity to crystal defects is captured clearly by the plots in Fig. 7. Fig 7(a) shows the temperature dependence of the electron-phonon, electron-electron and electron-impurity/defect scattering terms. Above ~600 K the electron-phonon and electron-electron scattering terms dominate. When defect removal due to annealing is considered, the electron-impurity scattering term is almost an order of magnitude lower at 1000 K than the self-scattering and electron-phonon scattering contributions.

According to the kinetic theory model used, the rate of change of the thermal diffusivity with respect to the defect density is given by

$$\frac{\partial \alpha}{\partial c_v} = -\frac{\alpha^2 3 C_P}{C_e v_F^2}\left[\frac{1}{\tau_{e,i}} + \frac{1}{\tau_{e,v}} - \frac{1}{\tau_e}\right]. \tag{9}$$

Fig. 7(b) shows a plot of eqn. 9 for pure tungsten from RT to 1073 K. The rate of change of the thermal diffusivity with respect to the defect density is negative, as expected. Its magnitude decreases with increasing temperature, showing a reduced sensitivity of the thermal diffusivity to defects at higher temperatures.

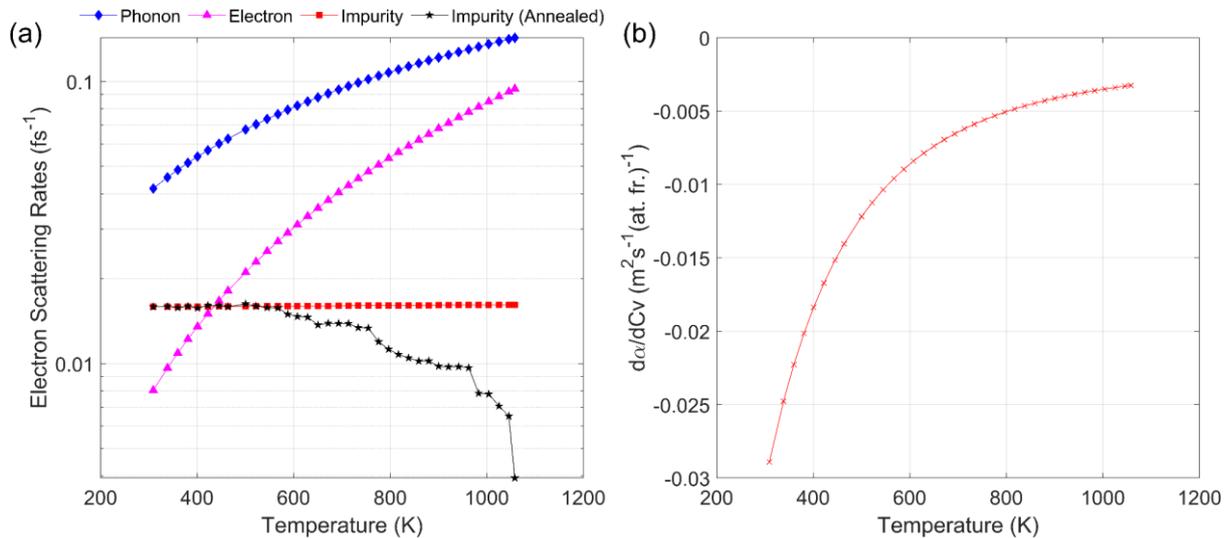

Figure 7. (a) Temperature dependence of the different electron scattering terms. The electron-electron and phonon-electron scattering rates are the same, irrespective of the level of damage in the sample. Given in red is the electron-impurity scattering rate assuming constant defect density with temperature, assuming the starting value of the $3.2 \times 10^{-1}$ dpa sample. Given in black is the electron-impurity scattering term when defects are removed through annealing in the $3.2 \times 10^{-1}$ dpa sample. (b) rate of change of thermal diffusivity with respect to equivalent Frenkel pair density as a function of temperature from the model introduced in this study for pristine tungsten.

The unannealed defect densities in Fig. 6, at higher doses / defect densities, are slightly higher than those previously determined by RT TGS on the same samples [7]. This is due to the difference in the Lorenz ratio '$L$' (proportionality constant in the Weidemann-Franz law) used in the current study. Generally for metals the Lorenz ratio takes the theoretical value (Sommerfeld value) of $2.44 \times 10^{-8}$ V$^2$ K$^{-2}$, which was the one used in [7] when computing the scattering rates. However for tungsten, experimentally determined values are appreciably higher, $3.0$-$3.4 \times 10^{-8}$ V$^2$ K$^{-2}$ at RT [71]. $3.2 \times 10^{-8}$ V$^2$ K$^{-2}$ was chosen as the Lorenz value for this work, also given in table 3.1. A higher Lorenz value gives lower electron-impurity scattering rates, resulting in higher defect density estimates, as seen in Fig. 6.

Also noticeable in Fig. 6 is that, with increasing dose, the defect density after the 473 K anneal decreases, implying some defect removal even before the temperature for mono-vacancy mobility is reached. Interestingly, this effect only becomes significant at the higher doses. The second stage of defect removal [40], where interstitials are released from traps can account for this. Studies have shown that at higher doses there are larger defect structures, e.g. dislocation networks, that can act as traps for interstitials [40,47]. The damage population at lower doses, on the other hand, is dominated by point defects and small defect clusters [24,74]. Thus in higher dose samples more trapped interstitials are present that are released between RT-500 K, giving a slight recovery of thermal diffusivity. Furthermore, the 873 K and 1073 K anneals also cause an appreciable decrease in defect density in the higher dose samples. This would be consistent with the onset of mobility for larger defects, such as di-vacancies and vacancy-impurity complexes [34,37].

## 3.3 Transmission electron microscopy

Fig. 8 shows TEM micrographs of FIB lift-outs from the 3.2 x 10$^{-2}$ dpa sample after the 1073 K annealing cycle (*a, b* and *c*) and the as-implanted 3.2 x 10$^{-3}$ dpa sample (*d,e* and *f*). These two cases were chosen as they both exhibit a similar thermal diffusivity of $\sim 5.0 \times 10^{-5}\, m^2 s^{-1}$ (Fig. 4) and mark the apparent transition from low dose to high dose behaviour. From Figs. 8(c) and 8(f) we see that there is a population of evenly spread defects in the unimplanted bulk. These are likely to be damage caused by the FIB beams used for sample manufacture (5 and 30 keV Ga ions) that have energy greater than the displacement threshold energy of tungsten (~68eV) [46]. Similar damage effects have been observed previously in FIB lift-outs of self-ion implanted tungsten [75]. Hence, in the micrographs taken at the interface between the implanted and unimplanted regions (Figs. 8(b) and 8(e)), we see some crystal damage in the unimplanted region also.

In the implanted region of the annealed 3.2 x 10$^{-2}$ dpa sample (Fig. 8(b)) we see larger defect structures unlike in the as-implanted 3.2 x 10$^{-3}$ dpa sample (Fig. 8(e)). The annealing is believed to have caused defect removal, as well as the growth of these larger defect structures. Since the FIB lift-outs were carried out after the implantation and annealing, we do notice smaller defects from the FIB damage in the 3.2 x 10$^{-2}$ dpa implanted region. However this effect is seen to be less prominent, perhaps because the larger defects act as a defect sinks during the FIB process. In the 3.2 x 10$^{-3}$ dpa sample however, there is no clear difference between the implanted and unimplanted region. This suggests that in this sample, there aren't any large defect structures at the end of the implantation stage.

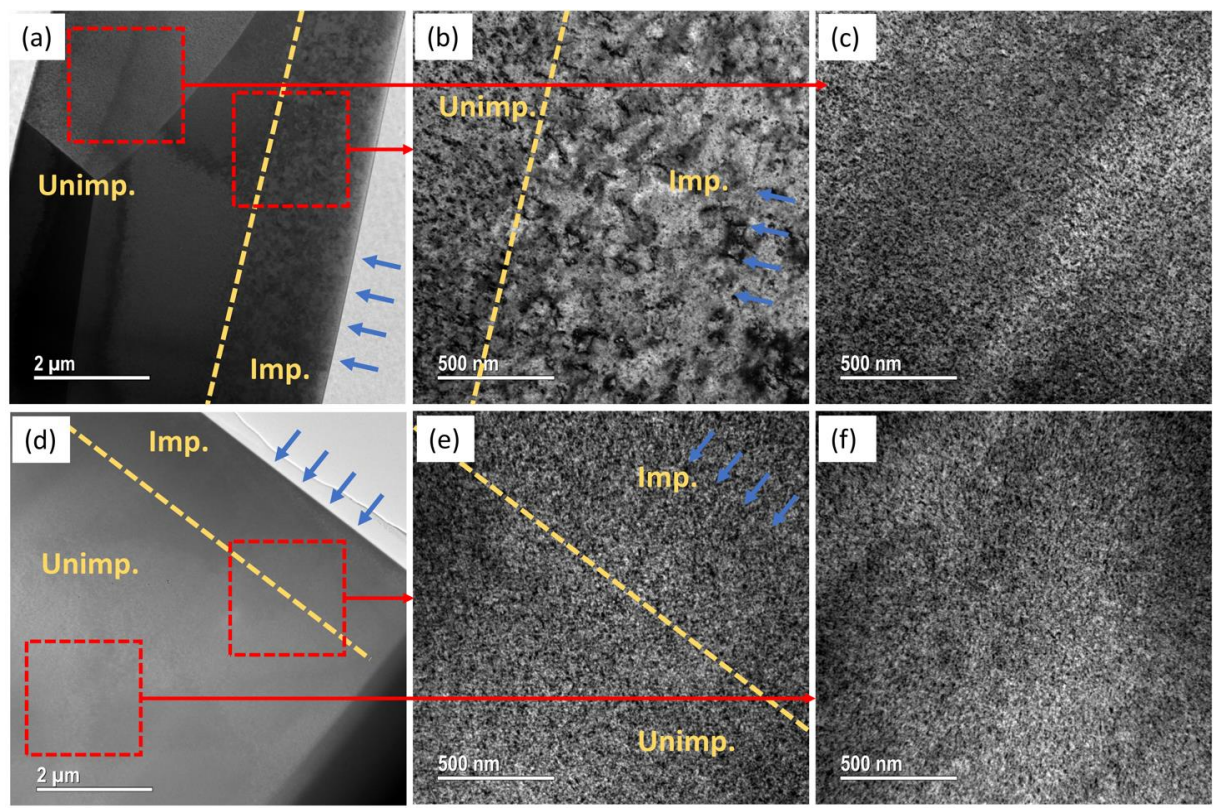

Figure 8. TEM micrographs of the 3.2 x 10$^{-2}$ dpa sample after the 1073 K annealing cycle (a) overview, (b) near surface layer with implanted and unimplanted regions, and (c) unimplanted bulk. 3.2 x 10$^{-3}$ dpa as-implanted sample TEM micrographs show (d) overview, (e) near surface layer with implanted and unimplanted regions, and (f) unimplanted bulk. The interface between the implanted and unimplanted regions is marked by the yellow dashed line. Blue arrows indicate the implantation direction and sample surface.

## 4 Discussion

An interesting point concerns the increase in thermal diffusivity with temperature observed in the higher dose samples (3.2 x 10$^{-1}$ dpa and 3.2 dpa), at temperatures below the threshold for vacancy mobility (see the 473 K annealing cycle in Fig. 3). This effect is not one of defect removal as the thermal diffusivity curve upon cooling retraces the heating one. The expected thermal diffusivity when there is no annealing taking place in the 3.2 dpa sample is plotted in Fig. 3(e) (black crosses) assuming the initial defect density. At lower temperatures it clearly shows an increase in thermal diffusivity with temperature. A similar behaviour has been observed in tungsten alloyed with rhenium, where, unlike the crystal defects, the rhenium "impurities" are not removed by annealing [9,21]. This effect is the result of several competing phenomena: The thermal conductivity is inversely proportional to the three term electronic scattering expression given in eqns. 2 and 3 in section 3.2. This varies with temperature in the functional form: $a + bT + cT^2$, where $T$ is the absolute temperature. The thermal diffusivity is dependent on the thermal conductivity, the electronic heat capacity that is linear in temperature ($gT$), and the phonon heat capacity. The dependence of the phonon heat capacity on temperature near the Debye temperature is weaker than a linear relationship, perhaps proportional to $T^{0.5}$, approaching a constant value as the Dulong-Petit limit is reached [64]. The phonon heat capacity is greater than the electronic heat capacity by approximately two orders of magnitude and thus dominates behaviour. Ultimately, this gives a functional dependence of thermal diffusivity on temperature with the functional form $1/(aT^{-0.5} + bT^{0.5} + cT^{1.5})$. Hence, for sufficiently large values of 'a', there can be a positive slope of the thermal diffusivity with temperature. Indeed this effect could be exhibited in any regime where the phonon heat capacity has a sub-linear temperature dependence. This is useful for our analysis, as we can look for when the actual annealing curves deviate from the "no annealing" curve (black crosses in Fig. 3(e)) as an indication of when defects begin to be removed. Since this deviation falls into the range of temperatures associated with the onset of vacancy mobility, we hypothesize that it is the vacancy movement that leads to this first large defect removal step in our work.

The microstructural changes inferred from the measured thermal diffusivity evolution can be further substantiated by considering the TEM observations as well as previous literature studies: For the lower dose samples (3.2 x 10$^{-4}$ and 3.2 x 10$^{-3}$ dpa), we observe near total recovery of the thermal diffusivity by the end of the 1073 K anneal. These low doses have not been covered in TEM studies, understandably, as we hypothesize that the defects would be mostly point defects and invisible in TEM. A resistivity recovery study, where the resistivity of low dose neutron irradiated tungsten was monitiored with annealing temperature showed complete recovery by ~700 K [38]. Interestingly, the Frenkel pair concentration in this study was 4.92 x 10$^{-4}$ at.fr., simmilar to the lowest dose samples in our study, which also showed complete defect removal.

Unlike the lower dose samples, the higher dose ones, 3.2 x 10$^{-2}$ dpa and above, undergo a partial recovery of ~70% by the end of the 1073 K anneal. The low temperature anneal to 473 K showed recovery levels of ~5%, attributed to the release of defects from traps. The 673 K anneal had the largest defect recovery (~50%), showing significant annealing in the 473 K-673 K temperature range. This was seen to correlate well with PALS results that indicate mono-vacancy mobility in this range of temperature [39,73]. The hypothesis is also supported by creation-relaxation algorithm MD studies that predict a supersaturation of vacancies at higher doses [27,76].

After annealing the high dose samples to 873 K and 1073 K, the pristine value of thermal diffusivity is still not recovered. Partial recovery of the thermal diffusivity suggests that there is a remaining defect population even after annealing to 1073 K. The TEM results confirmed this hypothesis as large defect structures are clearly visible in the implanted region of the 3.2 x 10$^{-2}$ dpa sample after the 1073 K annealing cycle. No such large defect clusters are visible in the 3.2 x 10$^{-3}$ dpa unannealed sample.

It is interesting to compare our findings and predicted defect number densities to previous observations. Due to the time consuming nature of performing quantitative analysis of defect populations using TEM, few studies have been conducted for ion-implanted tungsten [24,34]. The reported defect loop densities were combined with the loop size distribution data to convert TEM data to equivalent Frenkel pair density for comparison with our results (see Fig. 9). The procedure for this is explained in detail in [7]. In Fig. 9 we see that the TEM defect densities from Yi et. al [24] are an order of magnitude lower than the TGS estimates. It is known that TEM lacks sensitivity to defects smaller than ~1.5 nm in diameter [7,28,43,77]. At low irradiation doses, the defect size distribution follows an inverse power law, giving large numbers of small defects [23,25]. Hence for a quantitative comparison (magnitudes and trends) of the TGS defect density predictions in this study, data for smaller defects, usually obtained from MD [7], is needed. The room temperature data from [24] has been supplemented with MD data [23], to give the blue square markers in Fig. 9. This MD+TEM data agrees well with the as-implanted RT defect predictions from TGS. Comparing the annealing effects directly is not possible as no high temperature MD data is available. We note that there is a decrease of ~50% in the defect densities between the RT and 773 K TEM data from [24], which is simillar to the change observed with TGS, between RT and 673 K. This is an interesting finding since the TEM work in [24] was done with insitu implantation at the relevant elevated temperatures, rather than the post-irradiation annealing as in this work. The data from [34] in Fig. 9 is significantly lower than the other TEM studies. The reason is that in [34] only defects with diameters larger than 4-5 nm were considered, whereas [24] considered defects down to ~1.5 nm.

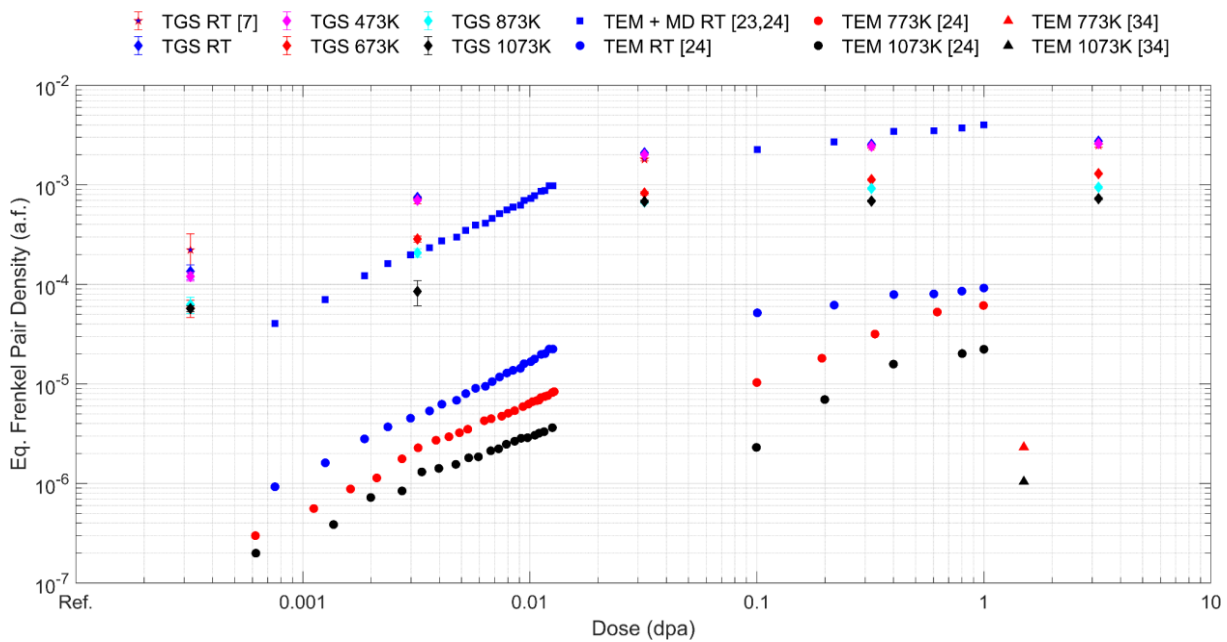

Figure 9. Comparison of TGS predicted defect densities with TEM [24,34] and MD [23] estimates. The TGS data are the estimates at RT after the relevant anneals whereas the TEM data are insitu.

A recent study using X-ray diffraction and atomic scale simulations showed that ion-implanted tungsten undergoes a complex sequence of structural changes as a function of damage dose [47]. Initially up to 0.01 dpa, there is an accumulation of Frenkel pairs as a result of displacement cascades.

At higher doses, beyond 0.01 dpa, damage starts to overlap and interact, resulting in the clustering of self-interstitials while vacancies are immobile (<600 K). As self-interstitials continue to cluster they eventually form new atomic planes. This is an effective mechanism for the removal of self-interstitials from the system without need for free surfaces. Since vacancies are immobile below ~600 K, they stay in the system. The result is an excess of finely-dispersed vacancies as well as large scale defect structures that form at high doses. Recently it has become possible to model this whole evolution using massively parallel MD simulations [27,47,76]. In [76] the thermal diffusivity degradation associated with the MD-predicted microstructure is computed using a kinetic theory model similar to that used in this study and [9]. The findings agreed well with the measured thermal diffusivity as a function of dose in the RT implanted tungsten [7]. This further confirms that small defects, particularly of vacancy type, that are too small to be discerned by TEM play a substantial role in the degradation of thermal diffusivity.

Thus, by combining TGS thermal diffusivity measurements, defect density predictions, TEM observations, lattice strain measurements, atomistic simulations and PALS measurements, a complete picture of the underlying irradiation defect evolution in tungsten and its effect on material properties begins to emerge.

# 5 Conclusions

A comprehensive study has been carried out on the thermal diffusivity degradation in ion-implanted tungsten, and its evolution with temperature, from RT to 1073 K, by transient grating spectroscopy with insitu heating. A simple kinetic theory model was used to estimate defect densities and transmission electron microscopy was used to verify the inferred microstructural evolution. The following key points are noted:

- The insitu heating stage designed and implemented in this study allows TGS measurements of thermal diffusivity and SAW velocity in thin ion-implanted layers from RT-1073 K.
- The temperature evolution of thermal diffusivity deviates from that expected for a static microstructure, indicating significant changes of the defect population in this temperature range.
- A kinetic theory model was introduced that allows the interpretation of thermal diffusivity in terms of evolving defect population. By making a relative comparison between implanted samples and pristine material from the same batch, rather than the book value, more accurate interpretation was possible.
- The largest removal of defects occurs in the temperature interval between 500-600 K. This coincides with the temperature window at which mono-vacancies become mobile. This is also consistent with a previous hypothesis that there is a large population of vacancies in RT implanted tungsten.
- The thermal diffusivity recovery associated with the vacancy mobility step is up to 50% of the RT thermal diffusivity. This shows that vacancies and small vacancy clusters are major contributors to irradiation-induced degradation of thermal transport.
- The estimated changes in defect populations are consistent with ex-situ TEM observations in this study and in the literature.
- The results obtained show that beyond ~1000 K, implantation induced defects have relatively little effect on thermal diffusivity. This is good news for fusion reactors, where some parts of the armour are expected to operate beyond this temperature.

# 6 Acknowledgements

We are grateful to Sergei Dudarev and Daniel Mason from the Culham Centre for Fusion Energy for fruitful discussions and constructive criticism. We thank Artem Lunev from the Culham Centre for Fusion Energy and Sam Humphry-Baker from Imperial College London for their help with heater calibration, and Roger Reed, Yuanbo Tang and Govind Gour, all from the University of Oxford, for their assistance in manufacturing thermocouples. This research was funded by the European Research Council (ERC) under the European Union's Horizon 2020 Research and Innovation Programme (StG AtoFun, grant No. 714697). The views and opinions expressed herein do not necessarily reflect those of the European Commission. C.A.D. acknowledges support from the Center for Thermal Energy Transport under Irradiation (TETI), an Energy Frontier Research Center funded by the US Department of Energy, Office of Science, Office of Basic Energy Sciences.

# 7 Appendix A

## Supplementary Figures

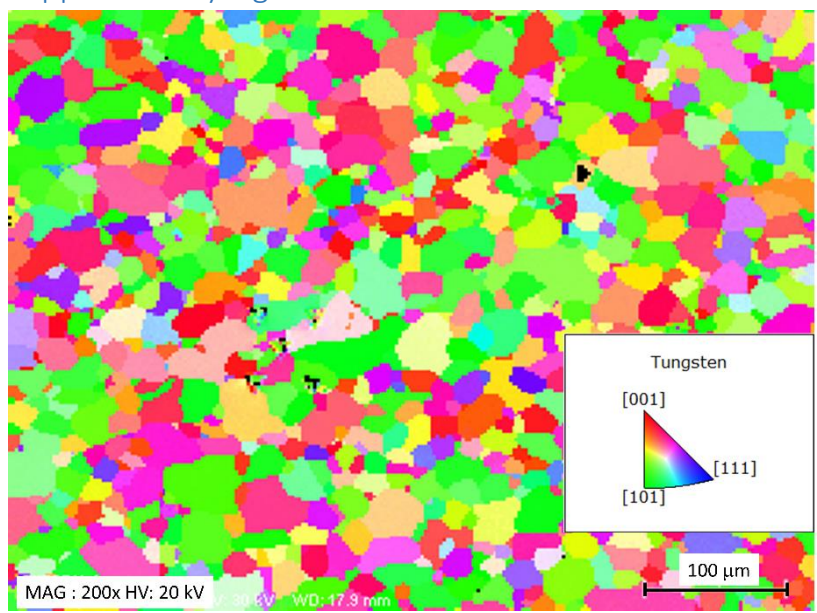

Figure A.1 EBSD map of the $3.2 \times 10^{-2}$ dpa sample after preparation annealing and polishing, obtained using a Zeiss MERLIN FEG SEM.

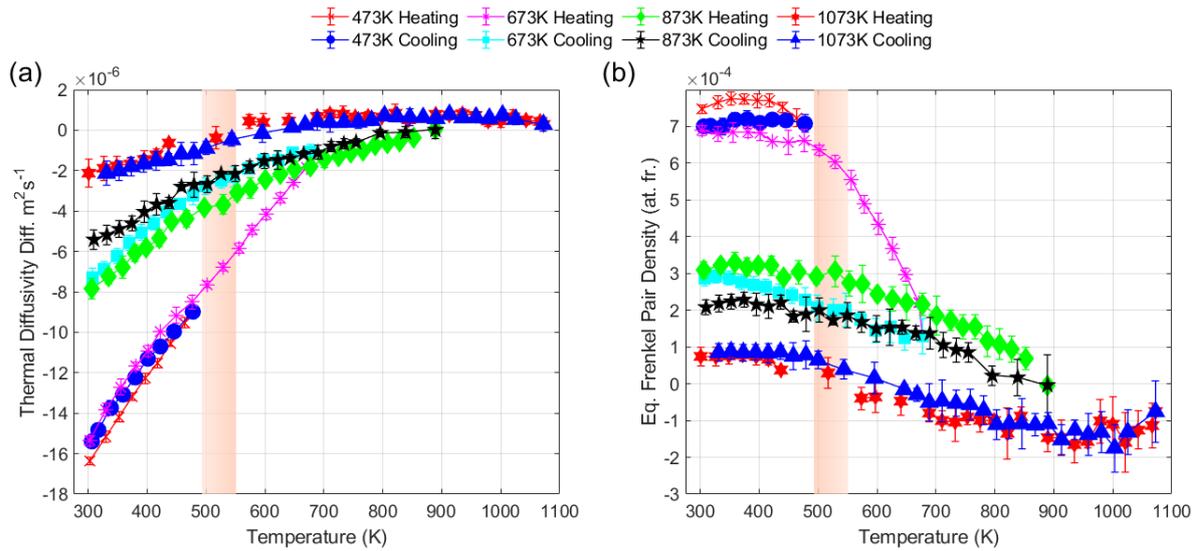

Figure A.2 Thermal diffusivity degradation (a) and estimated equivalent Frenkel pair density (b) for the 3.2 x $10^{-4}$ dpa sample. The shaded regions indicate the temperature range of mono-vacancy mobility [57].

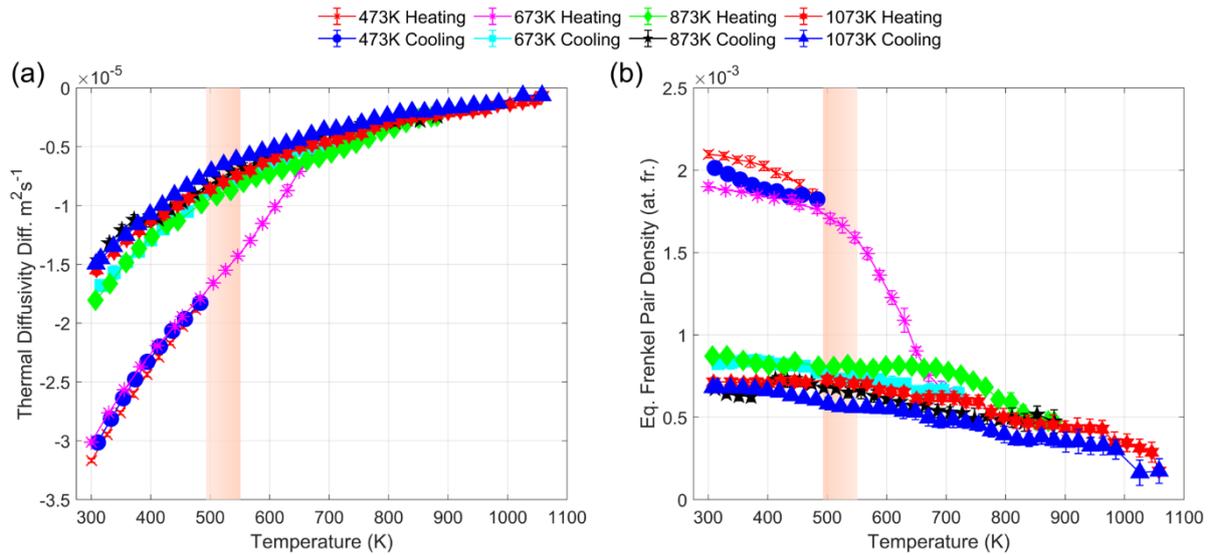

Figure A.3 Thermal diffusivity degradation (a) and estimated equivalent Frenkel pair density (b) for the 3.2 x $10^{-2}$ dpa sample. The shaded regions indicate the temperature range of mono-vacancy mobility [57].

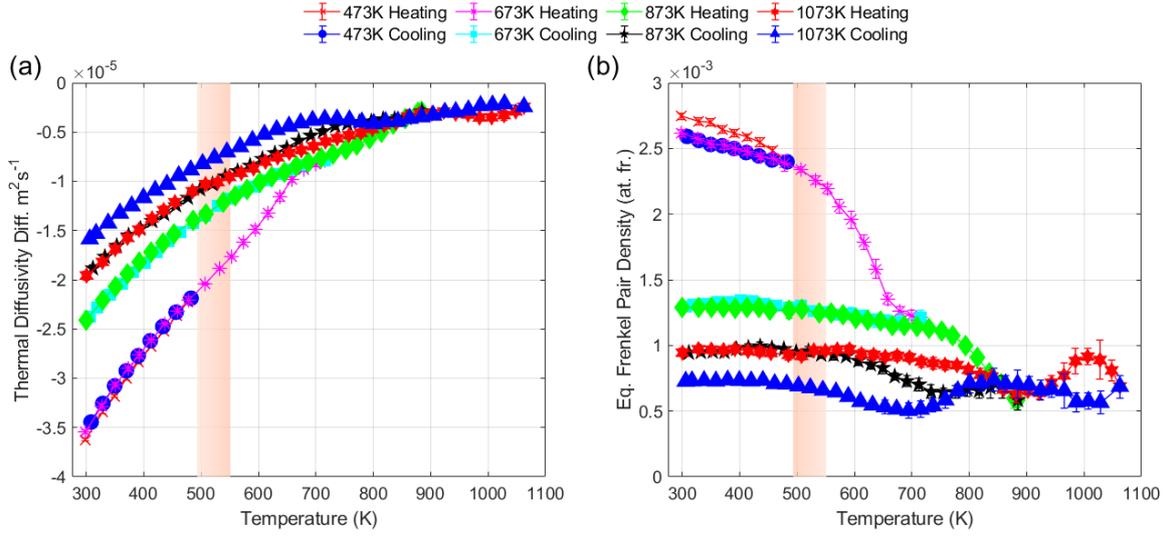

Figure A.4 Thermal diffusivity degradation (a) and estimated equivalent Frenkel pair density (b) for the 3.2 dpa sample. The shaded regions indicate the temperature range of mono-vacancy mobility [57].

# 8 Appendix B
## Fitting formulae and error analysis

The TGS traces were fitted with the following equation [9] using MATLAB [60]:

$$I = A\,\mathrm{erfc}(q\sqrt{\alpha t}) + C\sin(2\pi f t + E)\exp(-t/F) + G, \quad (B.1)$$

where A, C, E, F, G, $\alpha$ and $f$ are free parameters determined by the fitting. $\alpha$ is the thermal diffusivity, $f$ is the SAW frequency, and $t$ is time. Fits with a range of starting times covering one full SAW oscillation period were used to reduce the uncertainty associated with the exact starting point of the fit. The reported value for each measurement is the average over the range of different fit starting points.

The experimental uncertainty in the measurement of the thermal diffusivity $\alpha$ creates an uncertainty in the obtained value for the eq. Frenkel pair density. The eq. Frenkel pair density, $c_{FP}$, is given by

$$c_{FP} = c_v = c_i = \left[\frac{\Delta\alpha}{[\tau_e(\sigma'_i + \sigma'_v) - 2]\left[\frac{\tau_e C_e v_F^2}{3C_P} - \Delta\alpha\right]}\right]. \quad (B.2)$$

The uncertainty in $c_{FP}$ is derived as follows:

$$\delta(c_{FP}) = \left|\frac{\partial c_{FP}}{\partial \alpha}\right|\delta(\Delta\alpha),$$

$$\delta(c_{FP}) = \left| \frac{\frac{\tau_e C_e v_F^2}{3C_P}}{[\tau_e(\sigma_i' + \sigma_v') - 2]\left[\frac{\tau_e C_e v_F^2}{3C_P} - \Delta\alpha\right]^2} \right| \delta(\Delta\alpha), \quad \text{(B.3)}$$

where $\delta(\Delta\alpha)$ is the uncertainty in the thermal diffusivity difference $\Delta\alpha$ which is the thermal diffusivity difference between the measured value and that measured for a pristine sample. $\sigma_i' = 1/\tau_{e,i}$ and $\sigma_v' = 1/\tau_{e,v}$ are the adjusted scattering rates for interstitials and vacancies respectively. $\rho$ is the mass density [61], $C_P$ is the specific heat capacity [61], $C_e$ is the electronic heat capacity [70], $v_F$ is the Fermi velocity [70] and $\tau_e$ is the electron scattering time. It should be noted that since $\Delta\alpha = \alpha_{pure} - \alpha_{implanted}$, where $\alpha_{pure}$ is the thermal diffusivity measured for a pristine sample, $\delta(\Delta\alpha) = \sqrt{2}\delta(\alpha)$, where $\delta(\alpha)$ is the uncertainty of the TGS thermal diffusivity measurement.